\documentclass[twocolumn]{aastex631}
\usepackage{bm}
\everymath{\displaystyle}

\shorttitle{Supernova polarization signals from interactions with the 
dense CSM disk
}
\shortauthors{Wen et al.}
\usepackage{natbib}
\usepackage{amsmath}
\usepackage{CJK}
\usepackage[whole]{bxcjkjatype} 
\usepackage{CJKutf8}
\begin{document}

\title{Supernova Polarization Signals From the Interaction with a Dense Circumstellar Disk
}

\author[0009-0007-3401-7133]{Xudong Wen}
\affiliation{School of Physics and Astronomy, Beijing Normal University, Beijing 100875, China}
\affiliation{Institute for Frontier in Astronomy and Astrophysics, Beijing Normal University, Beijing 102206, China}
\email{wenxudong@mail.bnu.edu.cn}

\author[0000-0003-2516-6288]{He Gao*}
\affiliation{School of Physics and Astronomy, Beijing Normal University, Beijing 100875, China}
\affiliation{Institute for Frontier in Astronomy and Astrophysics, Beijing Normal University, Beijing 102206, China}
\email{gaohe@bnu.edu.cn}

\author[0000-0002-6535-8500]{Yi Yang
\begin{CJK*}{UTF8}{gbsn}
(杨轶)
\end{CJK*}}
\affiliation{Physics Department, Tsinghua University, Beijing, 100084, China}

\author[0000-0002-8708-0597]{Liangduan Liu}
\affiliation{Institute of Astrophysics, Central China Normal University, Wuhan 430079, People's Republic of China}
\affiliation{Key Laboratory of Quark and Lepton Physics (Central China Normal University), Ministry of Education, Wuhan 430079, People's Republic of China}

\author[0000-0002-9165-8312]{Shunke Ai}
\affiliation{Department of Astronomy, School of Physics and Technology, Wuhan University, Wuhan 430072, China}

\author[0000-0003-1904-0574]{Zongkai Peng}
\affiliation{School of Physics and Astronomy, Beijing Normal University, Beijing 100875, China}
\affiliation{Institute for Frontier in Astronomy and Astrophysics, Beijing Normal University, Beijing 102206, China}

\begin{abstract}
There is increasing evidence that massive stars may exhibit an enhanced mass loss shortly before their termination explosion. Some of them also indicate the enhancement of their circumstellar matter (CSM) is not spherically symmetric.
Supernova (SN) interacting with aspherical CSM could induce special polarization signals from multiple radiation components that deviate from spherical symmetry.
We investigate the time-evolution of the continuum polarization induced by the SN ejecta interacting with a disk/torus-like CSM.
Our calculation suggests that the interaction between the SN ejecta and an immediate disk-like CSM with a thin, homogenous density structure would produce a high continuum polarization, which may reach a peak level of $\sim$15\%. The interplay between the evolving geometry of the emitting regions and the time-variant flux ratio between the polar ejecta and the equatorial CSM interaction may produce a double-peaked feature in the polarization time sequence.
A similar trend of the time evolution
of the polarization is also found for a radially extended CSM disk that exhibits a wind-like density structure, with an overall relatively  lower level of continuum polarization ($<2.5\%$) during the interaction process.
We also identify a non-uniform temperature distribution along the radial direction of the CSM disk, which yields a strong wavelength dependence of the continuum polarization.
These signatures provide a unique geometric diagnostic to explore the interaction process and the associated extreme mass loss of the progenitors of interacting transients.

\end{abstract}


\section{Introduction} \label{sec:intro}

Modern high-cadence wide-field optical surveys (e.g., \citealp{Bellm2019, Graham2014}) are opening up a new era in our understanding of supernovae (SNe) and various types of transients by extending the boundaries of the transient phase space. 
Lightcurves and follow-up observations spanning from the first few hours to years or even decades after the transient explosion enable systematic search and comprehensive characterization of any signatures of the expanding ejecta interacting with the circumstellar matter (CSM). 
CSM in the immediate vicinity of SNe would be ionized by the initial shock breakout flash and manifest as bright and narrow emission lines of elements at rather high ionization states. Spectroscopy before such short-lived `flash' features swept away by the ejecta reveals that Type II and Ib SNe are commonly surrounded by a shell of matter from enhanced pre-explosion mass loss \cite{Quimby2007, Gal-Yam2014, Yaron2017, Bruch2021, Bruch2023}. 
Additionally, core-collapse (CC) SNe may also embedded in a substantial amount of dense CSM that extends to a rather large distance from the progenitor. They would sine for years owing to radiation from the expanding ejecta that interacts with the H-rich CSM, e.g., the Type IIn SNe. Moreover, 
superluminous supernovae (SLSNe) \citep{Inserra2018} and fast optical transients \citep{Smith2007,Ho2019,Ofek2010,Moriya2013} reveal occasional extreme mass loss of massive stars during the final evolutionary phase. 
It is currently believed that stellar instability \citep{Yoon2010,Quataert2012,Fuller2017} and binary interaction \citep{Chevalier12,Soker13}, including the Roche-lobe overflow (RLOF \citealp{Mohamed2007}), which leads to an elevated mass-loss in a binary system compared to that of the homogeneous stellar wind, are potential mechanisms causing the formation of CSM. However, the origin of mass loss of massive stars during the final evolutionary phase remains unclear, posing one of the biggest challenges in stellar evolution theory.

Substantial effort has gone into investigating the ejecta-CSM interaction process under a spherically-symmetric framework
\citep{Chev1982,Chevalier1994,Moriya2013,Dessart2015,Chevalier2017,liu2017,Margutti2022}, including numerical and analytical work to predict the lightcurves and spectra from the interaction process \citep{Chevalier2011,Ginzburg2012,Morozova2017,Suzuki2020,Metzger2022}, as well as modelling to explain specific events \citep{Ofek2010,Chatzopoulos13,Moriya2013}.
However, these models become inapplicable when the geometry of the CSM deviates from spherical symmetry, which could be described more properly with multi-dimensional calculations.
%A significant but highly uncertain effect results from the deviation from spherical symmetry, which could be described more properly with multi-dimensional calculations.
In the binary system scenarios where massive stars are prevalent, there could be equatorially-confined CSM disks formed by mass transfer and common envelope evolution \citep{Chevalier12,Pejcha2016,Metzger2017}. 
Such an intensive CSM enrichment process towards the orbital plane takes place as the companion fills its Roche lobe. 
For example, the early-time optical and X-ray properties and polarization evolution of AT2018cow can be attributed to the ejecta interacting with CSM confined in a compact and surrounding disk \citep{Margutti2019, Maund2023}. %The multiple peaks in the slowly declining light curves of iPTF14hls in $\sim$two years after the explosion are likely to be caused by the ejecta interacting with episodes of strong pre-explosion eruptions \citep{Wang2022}. 
The strong early polarization observed in some Type II SNe originates from the asymmetry of the disk-like CSM \cite{Bilinski2023}.
Some of the H-poor superluminous supernovae (SLSNe) present evidence of both H-rich \citep{Yan2017} and H-poor CSM \citep{Pursiainen2022}, the latter may also exhibit a high degree of asphericity as indicated by its high continuum polarization.

Deviations from spherical symmetry of the CSM can introduce high uncertainty in the observable properties of interacting transient sources. 
In particular, the lightcurve shape and the emission line profile could be highly dependent on the viewing angle. Major departures from spherical symmetry could also lead to an inadequate estimate of the bolometric luminosity and the total mass content of the CSM (see, e.g., \citealp{Yang2023}).
Despite numerical \citep{Nagao2020, Uno2023} and hydrodynamic calculations \citep{McDowell2018, Kurfurst2019} were carried out to characterize the photometric and polarimetry properties of the SN ejecta interacting with an aspherical CSM,
there is a lack of multi-dimensional exploration using polarization signals, in particular, the polarization evolution throughout the process that the ejecta engulfing an aspherical CSM distribution.
One of the simplest representations of asphericity yields the CSM centered on the SN and exhibits a density concentration towards a circular disk. This configuration possesses a single axial symmetry, which has been inferred from the well-defined dominant axis on the spectropolarimetric $Q-U$ plane for various cases. Such a disk-like CSM can be naturally expected as a consequence of binarity (e.g., \citealp{Vasylyev2023, Vasylyev2024, Mauerhan2024}).

Polarimetry provides a powerful diagnostics of the SN explosion geometry \citep{Shapiro82,Hoflich91,Kasen03,Wang08}.
As the SN ejecta expands and interacts with the CSM disk, additional radiation emitted from the interaction zone would be integrated into the observed bolometric signal of the SN.
\cite{Wen2023} proposed that two-component radiation would produce an observable polarization signal that is dependent on the viewing angle and the distribution of the emitting zone on the photosphere. The process of interacting with a CSM disk may induce a polarized signal similar to that produced by the large-scale inhomogeneous emission injection through multiple radiation components.

In this paper, we investigate the polarization properties of SN interacting with a dense disk-like
CSM, which could be used as a probe to explore the CSM geometry. We use a 3-dimensional (3D) Monte Carlo polarization simulation code (MCPSC) to calculate the lightcurves, spectra,
and polarization signals generated from SN ejecta and CSM interactions, where a dense disk-like CSM is centered at the progenitor star. We investigate the polarization and its temporal evolution for different CSM disk structures, namely across a range of opening angles of the CSM measured from the center of the exploding progenitor star.
We also examine the viewing-angle dependence of various observables.
Finally, we discuss the feasibility of
using wavelength-dependent polarization signals as a new method to study CSM geometry and decipher the SN pre-explosion mass history.

\section{Methods} \label{sec:Methods}

\begin{figure*}[tbph]
\begin{center}
\includegraphics[width=0.85\textwidth,angle=0]{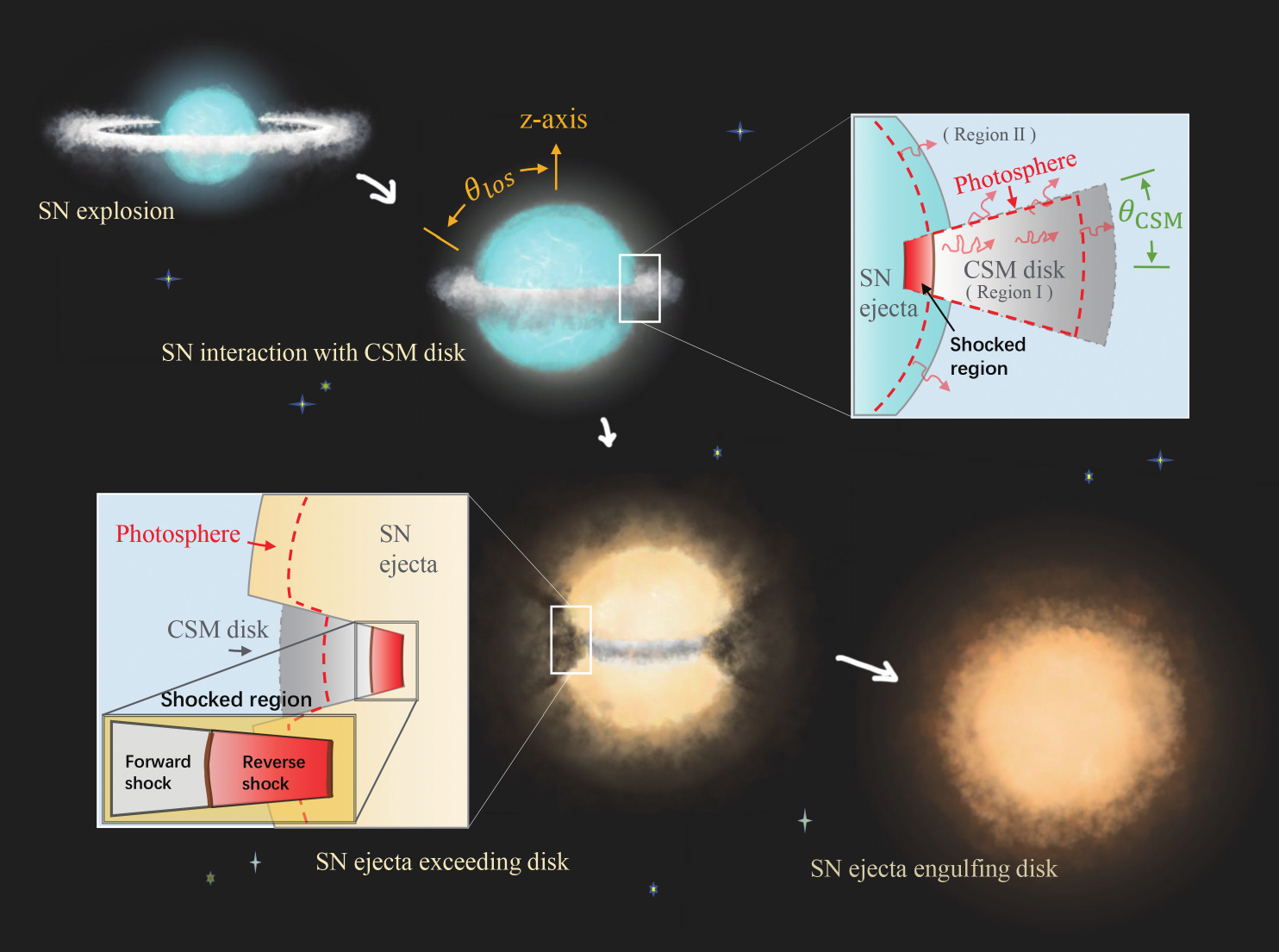}
\end{center}
\caption{Schematic illustration of the SN ejecta interacting with a dense, torus-like CSM concentrated to an arbitrarily defined equatorial plane.
When the interaction takes place with a half-tension angle $\theta$, 
the global geometry is divided into the interacting CSM disk (region I) and the expanding ejecta along the poles (region II). 
And the ejecta freely expanding towards the regions perpendicular to the plane of the CSM torus.
As the ejecta continues to expand, it will overrun the the CSM disk and may even engulf the CSM disk.
}
\label{fig:Fig_n1}
\end{figure*}

In this section, we describe the landscape landscape of the ejecta-CSM interaction, with the latter exhibits a density enhancement towards an arbitrarily defined plane, namely the equatorial plane.
When a massive star explodes as an SN, a total mass $M_{\rm ej}$ with an explosive energy $E_{\mathrm{SN}}$ is ejected. We assume that the ejecta undergoes a homologous expansion with velocity $v_{\rm{SN}}$ and interact with a CSM disk with inner boundary $R_{\text{CSM,in}}$. Two shocks are formed in the interaction region: a forward shock (FS) wave that traverses the CSM, and a reverse shock (RS) wave that sweeps over the supernova ejecta \citep{Chev1982,Chevalier1994}. This interaction converts kinetic energy into radiation, injecting additional energy into the equatorial plane.
The emitting zone can therefore be
divided into two parts, namely the interaction (Region \uppercase\expandafter{\romannumeral1}, $ \pi/2 - \theta_{\rm csm} < \theta < \pi/2 + \theta_{\rm csm}$)), and the polar (Region \uppercase\expandafter{\romannumeral2}, ($|\pi/2 - \theta| > \theta_{\rm csm}$) regions, where $\theta_{\rm csm}$ denotes the half-opening angle of the CSM disk measured from the equatorial direction (see, Figure~\ref{fig:Fig_n1}). The emission of the former zone is a combination of the ejecta-CSM interaction and the radioactive decay of Fe-group elements. The radiation of the latter region would be entirely dominated by radioactive decay.

In the case of CSM that extended to a large distance from the progenitor star (i.e., $\gtrsim10^{15}$\,cm), the early phase of the interaction takes place at the shocked region. The unshocked CSM exposed outside the SN photosphere thus forms a torus along the equatorial plane.
As the ejecta continues to expand, it will sweep across the entire CSM disk and expand radially outward, forming a shape characterized by elongated poles and a shorter equator. 
The interaction process terminates
when the shock wave sweeps across the outer radius of the CSM $R_{\text{CSM,out}}$. Eventually, as suggested by certain numerical simulation results, the ejecta may completely engulf the CSM as suggested by numerical calculations \citep{McDowell2018}.
During this process, the relative strengths of the light emitted from the two regions, as well as the non-spherical geometry of the CSM and SN ejecta, may give rise to observable polarisation signals. 

In order to describe the temporal evolution of the polarization properties, we conduct three-dimensional calculation of the polarization properties using the Monte Carlo Polarization Simulation Code (MCPSC, \citealp{Wen2023}). 
The code computes the flux and the 
polarization spectra by simulating the random walk of $N_{\rm{MC}}$ Monte Carlo (MC) photon packets through the propagating medium. There are $N_{\rm{\uppercase\expandafter{\romannumeral1}}}$ and $N_{\rm{\uppercase\expandafter{\romannumeral2}}}$ photon packets emitted from the given boundaries in Regions \uppercase\expandafter{\romannumeral1} and \uppercase\expandafter{\romannumeral2}, respectively, and satisfying
$N_{\rm{MC}} = N_{\rm{\uppercase\expandafter{\romannumeral1}}} + N_{\rm{\uppercase\expandafter{\romannumeral2}}}$. 
The computational domain is discretized into a $100 \times 100 \times 100$ Cartesian grid to configure the computing environment. 

\subsection{The Density Profile of the SN Ejecta}~\label{sec:ejecta_profile}
Following \cite{matzner99}, we adopt a broken power law to describe the radial density profile of the SN ejecta, i.e., 

\begin{equation}
  \rho_{\mathrm{ej}} (v, t) = \left\{ \begin{array}{ll}
    \zeta_{\rm{\rho}} \frac{M_{\mathrm{ej}}}{R_{\mathrm{tr}}^3} \left(
    \frac{r}{R_{\mathrm{tr}}} \right)^{- \delta}, & r < R_{\mathrm{tr}}\\
    \zeta_{\rm{\rho}} \frac{M_{\mathrm{ej}}}{R_{\mathrm{tr}}^3} \left(
    \frac{r}{R_{\mathrm{tr}}} \right)^{- n}, & r
    \geq R_{\mathrm{tr}}
  \end{array} \right.
\end{equation}
where the transition radius $R_{\text{tr}}$ of the ejecta is expressed as
\begin{equation}
  R_{\mathrm{tr}} = \zeta_{\rm{v}} \left( \frac{E_{\mathrm{SN}}}{M_{\mathrm{ej}}}
  \right)^{1 / 2} t.
\end{equation}
The coefficients can be obtained from the normalized continuity condition, namely
\begin{equation}
  \zeta_{\rm{\rho}} = \frac{(n - 3) (3 - \delta)}{4 \pi (n - \delta)}, \hspace{1em}
  \zeta_{\rm v} = \left[ \frac{2 (5 - \delta) (n - 5)}{(n - 3) (3 - \delta)}
  \right]^{1 / 2} .
\end{equation}
For a red supergiant progenitor, the typical values of the density power indices are $\delta = 1, n = 12$ \citep{matzner99}. 

\subsection{The Density Profile of the CSM}~\label{sec:csm_profile}
Following the prescription of \citep{Suzuki2019}, we describe the density profile of a torus-like CSM as'
\begin{equation}
\rho _{\text{CSM}}\left( r,\theta\right) =qr^{-s} \exp\bigg [-\left(\frac{r}{R_{\text{CSM,out}}}\right)^{p}-\left(\frac{\theta}{\theta_{\rm{CSM}}}\right)^{m}\bigg].
\label{rho}
\end{equation}
For the sake of simplicity,
we follow \cite{Nagao2020} and neglect the terms that smooth the edge density involving $R_{\text{CSM,out}}$ and $\theta_{\rm{csm}}$. The density profile thus yields
$\rho _{\text{CSM}}\left( r\right) =qr^{-s}$ along the radial direction.
The scaling factor $q$ can be determined as $q=\rho _{\text{CSM,in}
}R_{\text{CSM,in}}^{s}$, where $\rho _{\text{CSM,in}}$ denotes the density at the inner radius $R_{\text{CSM,in}}$ of the CSM.
The power-law index of the CSM density profile yields $s$, which $s=0$ and $s=2$ characterize a thin shell and a homogeneous wind, respectively \citep{Chatzopoulos12}.
The mass of the CSM disk ${M}_{\rm{disk}}$ can be written as
${M}_{\rm{disk}} = {M}_{\rm{CSM,sphere}} \sin{\theta_{\rm{csm}}} $, where ${M}_{\rm{CSM,sphere}}$ is the total mass of the spherical CSM.

\subsection{Shock dynamics and luminosity input}~\label{sec:forward_reverse}
The dynamical evolution of shock waves along the radial direction of the CSM disk follows the self-similar solutions found by \cite{Chev1982}. The radius of the forward ($R_{\mathrm{sh,FS}}$) and reverse shocks ($R_{\mathrm{sh,RS}}$) can be expressed as a function of time
, i.e.,
\begin{equation}
    \begin{aligned}
  R_{\mathrm{sh,FS}} = &R_{\text{CSM,in}} + \beta_{\rm F}\left[\frac{Ag^{n}}{q} \right]^{\frac{1}{n-s}}t^{\frac{n-3}{n-s}}\\ &(t_{i}<t<t_{i}+t_{\mathrm{FS}}),
  \end{aligned}
\end{equation}
and
\begin{equation}
    \begin{aligned}
  R_{\mathrm{sh,RS}} = &R_{\text{CSM,in}} + \beta_{\rm R}\left[\frac{Ag^{n}}{q} \right]^{\frac{1}{n-s}}t^{\frac{n-3}{n-s}}\\ &(t_{i}<t<t_{i}+t_{\mathrm{RS}}),
  \end{aligned}
\end{equation}
Where $\beta_{\rm{F}}$, $\beta _{\rm{R}}$, and $A$ are constants determined by the values of n and s, as detailed in Table 1 of \cite{Chev1982}. Here $t_{i}$ denotes the time when the SN eject start to interact with the CSM,
$t_{\mathrm{FS}}$ and $t_{\mathrm{RS} }$ represent the termination times of
the FS sweeping through all the optically thick CSM material and the RS sweeping through the SN ejecta, respectively.

Each photon packet is initialized with a specific energy and frequency, determined by the luminosity and temperature of the emitting region's photosphere \citep{Mazzali93,Lucy99}. In this work, the luminosity input $L_{\mathrm{Ni},\mathrm{\uppercase\expandafter{\romannumeral1}}/\mathrm{\uppercase\expandafter{\romannumeral2}}}$ in the SN ejecta from radioactive element decay for regions \uppercase\expandafter{\romannumeral1} and \uppercase\expandafter{\romannumeral2} can be expressed as
\begin{equation}
  L_{\mathrm{Ni},\mathrm{\uppercase\expandafter{\romannumeral1}}/\mathrm{\uppercase\expandafter{\romannumeral2}}} (t) = M_{\mathrm{Ni},\mathrm{\uppercase\expandafter{\romannumeral1}}/\mathrm{\uppercase\expandafter{\romannumeral2}}}[(\mathrm{\epsilon_{Ni}}-\mathrm{\epsilon_{Co}})e^{-t/\tau_{\mathrm{Ni}}}+\mathrm{\epsilon_{Co}}e^{-t/\tau_{\mathrm{Co}}}],
\end{equation} 
where $\mathrm{\epsilon_{Ni}} = 3.9\times10^{10} \mathrm{erg~g^{-1}~s^{-1}}$ and $\mathrm{\epsilon_{Co}} = 6.8\times10^{9} \mathrm{erg~g^{-1}~s^{-1}}$ are the heating rates of $^{56}$Ni and $^{56}$Co, respectively. $\tau_{\mathrm{Ni}}$ = 8.8 days and $\tau_{\mathrm{Co}}$ = 111.3 days are their decay timescales \citep{Valenti2008,Khatami2019}. 
By assuming $^{56}$Ni is evenly distributed throughout the ejecta, the total ejecta mass $M_{\mathrm{Ni}}$ that is dispersed in regions \uppercase\expandafter{\romannumeral1} and \uppercase\expandafter{\romannumeral2} give'
\begin{equation}
  M_{\mathrm{Ni},\mathrm{\uppercase\expandafter{\romannumeral1}}} = M_{\mathrm{Ni}}\sin(\theta_{\mathrm{csm}}),  \quad M_{\mathrm{Ni},\mathrm{\uppercase\expandafter{\romannumeral2}}} = M_{\mathrm{Ni}} - M_{\mathrm{Ni},\mathrm{\uppercase\expandafter{\romannumeral1}}}, 
\end{equation} 
$M_{\mathrm{Ni}}$ is the mass of all $^{56}$Ni contained in the whole ejecta.

Following \cite{Chatzopoulos12} and \cite{Wang2019}, we take the luminosity input from the forward ($L_{\rm{FS}}$) and reverse ($L_{\rm{RS}}$) shock waves as 
\begin{equation}
\begin{aligned}
L_{\rm{FS}}(t) = &\frac{2\pi }{(n-s)^{3}}\left( Ag^{n}\right) ^{\frac{5-s}{
n-s}}q^{\frac{n-5}{n-s}}(n-3)^{2}(n-5) \\
& \beta _{\rm{F}}^{5-s}t^{\frac{2n+6s-ns-15}{
n-s}} \quad (t_{i}<t<t_{i}+t_{\mathrm{FS}})
\end{aligned}
\end{equation}
and 
\begin{equation}
\begin{aligned}
L_{\rm{RS}}(t)=&\frac{2\pi g^{n}}{\beta _{\rm{R}}^{n-5}}\frac{n-5}{n-3}\left( 
\frac{3-s}{n-s}\right) ^{3}\left( \frac{q}{Ag^{n}}\right) ^{\frac{n-5}{n-s}%
}\\
&t^{\frac{2n+6s-ns-15}{n-s}} \quad (t_{i}<t<t_{i}+t_{\mathrm{RS}}).
\end{aligned}
\end{equation}
Adopting an approach similar to that of \cite{Nagao2020},
the total heating luminosity in the CSM disk produced from the interaction process gives
\begin{equation}
L_{\mathrm{CSM,disk}}(t)=[\epsilon_{\rm{FS}}L_{\rm{FS}}(t)+\epsilon_{\rm{RS}}L_{\rm{RS}}(t)]\sin{\theta_{\rm{csm}}},
\end{equation}
where the $\epsilon_{\rm{FS}}$ and $\epsilon_{\rm{RS}}$ are the corresponding conversion efficiency from kinetic energy to radiation for forward and reverse shock waves, respectively.

\subsection{The Integrated Emission of the SN}~\label{sec:emission}
In region $\mathrm{\uppercase\expandafter{\romannumeral1}}$, we consider a hybrid model which the luminosity input is provided by a combination of the the ejecta-CSM
interaction and the radioactive decay. 
When calculating the luminosity of the photosphere embedding the CSM disk, 
it is necessary to consider the photon diffusion time from the shock plane to different areas of the photosphere, leading to variations in local luminosity of the photosphere. Photons in the CSM disk that are exposed outside the supernova photosphere in region $\mathrm{\uppercase\expandafter{\romannumeral2}}$ are more likely to escape from the latitudinal photosphere of the CSM disk. This is because the latitudinal optical depth is relatively smaller compared to the radial optical depth.
We therefore focus on the luminosity produced from photospheric areas at different radii of the CSM disk, and neglect the slight effect of the latitudinal optical depth.
We trace the random walk process of $N_{\rm{\uppercase\expandafter{\romannumeral2}}}$ photon packets from forward shock wave surface to the CSM disk photosphere to determine the distribution of photon packets on the CSM disk photosphere.
In practice, the CSM torus has been
divided into \(N_{\rm{l}}\) layers along the radial direction from \(R_{\rm{sh,FS}}\) to \(R_{\rm{ph,CSM}}\). For each layer \(i\), we track the number of photon packets passing through (\(N_{\rm{p},i}\)) and the number of photon packets radiating out from that layer (\(N_{\rm{rad},i}\)).
The luminosity generated by the ejecta-CSM
interaction in the \(i\)-th layer can be expressed as \(L_{\text{disk}}\) multiplied by the contribution rate $N_{\rm{rad},i}/N_{\rm{\uppercase\expandafter{\romannumeral2}}}$ for that layer, namely 
\begin{equation}
\begin{aligned}
&L_{\text{CSM}, \text{i}}\left( t\right) = \frac{N_{\rm{rad},i}}{N_{\rm{\uppercase\expandafter{\romannumeral2}}}}  \frac{1}{t_{\text{dCSM,i}}}\exp \left[ -\frac{t%
}{t_{\text{dCSM,i}}}\right]\\
&\int_{0}^{t}\exp \left[ \frac{t^{\prime }}{t_{%
\text{dCSM,i}}}\right] L_{\text{CSM,disk}}\left( t^{\prime }\right)
dt^{\prime },
\label{Temp8}
\end{aligned}
\end{equation}%
The diffusion timescale \(t_{\text{dCSM,i}}\) through the i layers of the CSM disk can be expressed as a function of the radius ($R_{\text{disk,i}}$) of the \(i\)-th layer in the radial direction 
and the total disk mass ($M_{\text{disk,i}}$) enclosed within $R_{\text{disk,i}}$, i.e.,
\begin{equation}
t_{\text{dCSM,i}}= \frac{\kappa _{\text{CSM}}M_{\text{disk,i}}}{
\beta c R_{\text{disk,i}} \sin{\theta_{\rm{csm}}}},
\label{Temp9}
\end{equation}
where $\kappa_{\rm CSM}$ is the mean opacity in CSM. $\beta = 13.8$ is a constant for the density distribution of the ejecta. Similarly, the luminosity contribution to the \(i\)-th layer from the decay of radioactive elements in the ejecta from region \(\mathrm{\uppercase\expandafter{\romannumeral1}}\) (\(L_{\rm{rad,\uppercase\expandafter{\romannumeral1}},i}\)) can be expressed as 
\begin{equation}
\begin{aligned}
&L_{\rm{rad}\mathrm{\uppercase\expandafter{\romannumeral1}},\text{i}}\left( t\right) = \frac{N_{\rm{rad},i}}{N_{\rm{\uppercase\expandafter{\romannumeral2}}}}  \frac{1}{t_{\text{drad,i}}}\exp \left[ -\frac{t%
}{t_{\text{drad,i}}}\right]\\
&\int_{0}^{t}\exp \left[ \frac{t^{\prime }}{t_{%
\text{drad,i}}}\right] L_{\rm{Ni},\mathrm{\uppercase\expandafter{\romannumeral1}}}\left( t^{\prime }\right)
dt^{\prime },
\label{Temp10}
\end{aligned}
\end{equation}%
and the diffusion timescale $t_{\text{drad,i}}$ through the i layers of the CSM disk is
\begin{equation}
t_{\text{d,i}}= \frac{\kappa _{\text{CSM}}\left[ (M_{\text{disk,i}}/\sin{\theta_{\rm{csm}}})+ M_{\text{ej}}\right]}{
\beta cR_{\text{disk,i}}}  
\label{Temp11}
\end{equation}
The energy carried by
each photon packet radiated from the \(i\)-th layer of the photosphere can be written as
\begin{equation} 
\epsilon_{\rm{i}} = \frac{\left[L_{\text{CSM,i}}\left( t\right)+L_{\rm{rad}\mathrm{\uppercase\expandafter{\romannumeral1}},\text{i}}\left( t\right)\right]\Delta t}{N_{\rm{rad,i}}},
\label{Temp12}
\end{equation}
where $\Delta t$ is the duration of Monte Carlo process.
The temperature radiated from layer i in the CSM disk can be expressed in terms of
the luminosity radiated from layer i and the equivalent radial cross-sectional area $A_{\rm{disk,i}}$ , i.e.,
\begin{equation}
  T_{\rm{disk}} = \left[\frac{L_{\text{CSM,i}}\left( t\right)+L_{\rm{rad}\mathrm{\uppercase\expandafter{\romannumeral1}},\text{i}}\left( t\right)}{\sigma_{\rm SB} A_{\rm{disk,i}}}\right]^{1/4} .
  \label{Temp13}
\end{equation}
When the radius of the supernova photosphere exceeds the outer boundary of the CSM disk, the luminosity calculation for region $\mathrm{\uppercase\expandafter{\romannumeral1}}$ can be simplified by assuming that photons within the disk can only escape radially from the photosphere of the disk.

As for region $\mathrm{\uppercase\expandafter{\romannumeral2}}$, the luminosity is mainly contributed by the radioactive element decay in the expanding SN ejecta. Based on \cite{Arnett82}, the luminosity could be estimated to be 
\begin{eqnarray}
L_{\rm{rad}\mathrm{\uppercase\expandafter{\romannumeral2}}}(t)&=&\frac{2}{t_{\rm{m}}}e^{-\left(\frac{t^{2}}{t_{\rm{m}}^{2}}+\frac{2R_{0}t}{vt_{\rm{m}}^{2}}\right)}~
\int_{0}^{t} e^{\left(\frac{t'^{2}}{t_{\rm{m}}^{2}}+\frac{2R_{0}t'}{vt_{\rm{m}}^{2}}\right)}   \nonumber\\
     &&\times\left(\frac{R_{0}}{vt_{\rm{m}}}+\frac{t'}{t_{\rm{m}}}\right)L_{\mathrm{Ni},\mathrm{\uppercase\expandafter{\romannumeral2}}}(t')dt',
\label{equ:lum region 1}
\end{eqnarray}
where $R_{0}$ is the initial radius of the progenitor, and we follow \cite{Chatzopoulos12} to take the limit $R_{0}\rightarrow0$. 
The diffusion time scale $t_{\rm{m}}$ can be written as
\begin{eqnarray}
t_{\rm{m}}=\left(\frac{2\kappa_{\rm{ej}} M_{\rm ej}}{\beta vc}\right)^{1/2},
\label{equ:tau_m}
\end{eqnarray}
where $\kappa_{\rm{ej}}$ is the mean opacity to optical photons in SN ejecta.
The energy carried by each photon packet emitted from the photosphere in region $\mathrm{\uppercase\expandafter{\romannumeral2}}$ is $L_{\rm{rad}\mathrm{\uppercase\expandafter{\romannumeral2}}} \Delta t/N_{\rm{\mathrm{\uppercase\expandafter{\romannumeral2}}}}$. The photosphere temperature is
 \begin{equation}
  T_{\rm disk} = \left(\frac{L_{\rm{rad}\mathrm{\uppercase\expandafter{\romannumeral2}}}}{\sigma_{\rm SB} A_{\rm{ej}} }\right)^{1/4} ,
\end{equation}
where $A_{\rm{ej}}$ is the area of the photosphere in the region $\mathrm{\uppercase\expandafter{\romannumeral2}}$  supernova ejecta.

\subsection{The Integrated Polarization of the SN}~\label{sec:polarization}
Each photon packet carries polarization information which can be described by a useful convention, namely the Stokes vector $S = (I, Q, U, V)$, where $I$ gives
the total intensity, $Q$ and $U$ describe the
linear polarization, and $V$ specifies the state
of circular polarization. Considering that circular polarization has never been observed in SNe, also, in a scattering atmosphere without a magnetic field, the radiative transfer calculations for circular and linear polarization can be decoupled \citep{chandrasekhar1960}, we therefore neglect the $V$ component.
The polarization degree ($P$) and the position angle ($\chi$) can then be given in terms of the Stokes parameters, namely
\begin{equation}
P = \frac{\sqrt{Q^2+U^2}}{I} ,
\end{equation}
\begin{equation}
\label{chi}
\chi = \frac{1}{2} \tan^{-1}{\bigg(\frac{U}{Q}\bigg)} ~ .
\end{equation}
For the sake of convenience, for the case of
the orthogonal components of $U$ is balanced in the projection plane along the line-of-sight to the observer, the polarization degree can be simplified to $P = Q/I$.

Unpolarized photon packets are emitted from the photosphere with random propagation directions and propagate through the medium. Following the prescriptions in previous studies (e.g., \citealp{Code95, Mazzali93, Lucy99, Kasen03, Whitney2011, Bulla15}), photon packets with zero polarization were prepared at the emitting layer, whose initial traveling directions were set randomly. For a photon packet traveling through the medium, a certain probability of electron scattering is assigned. 
After the scattering occurs, the renewed polarization state and the propagation direction of the photon packet will then be set as the initial conditions. We remark that our calculations focus on the continuum polarization that arises from the deviation from spherical symmetry of an electron-scattering photosphere. We do not attempt to model the polarization induced by particular line opacities. 
The continuum polarization in the optical band is independent of wavelength as Thomson scattering is approximately a grey process at this frequency \citep{Bulla2017}.
We adopt a conventional Thomson electron scattering opacity value of $\kappa=$0.34 for both $\kappa_{\rm CSM}$ and $\kappa_{\rm{ej}}$ \citep{Nagao2020}. 
When the photon packet reaches the outer boundary of the computational domain, the propagation process ends, and the information of the photon packet at that time is recorded. 
The observer frame was set at 100 Mpc from the SN center. The final collection of the photon packets is carried out within each $\frac{1}{16}\times180^{\circ}$ latitudinal bins.

\begin{figure*}[tbph]
\begin{center}
\includegraphics[width=1\textwidth,angle=0]{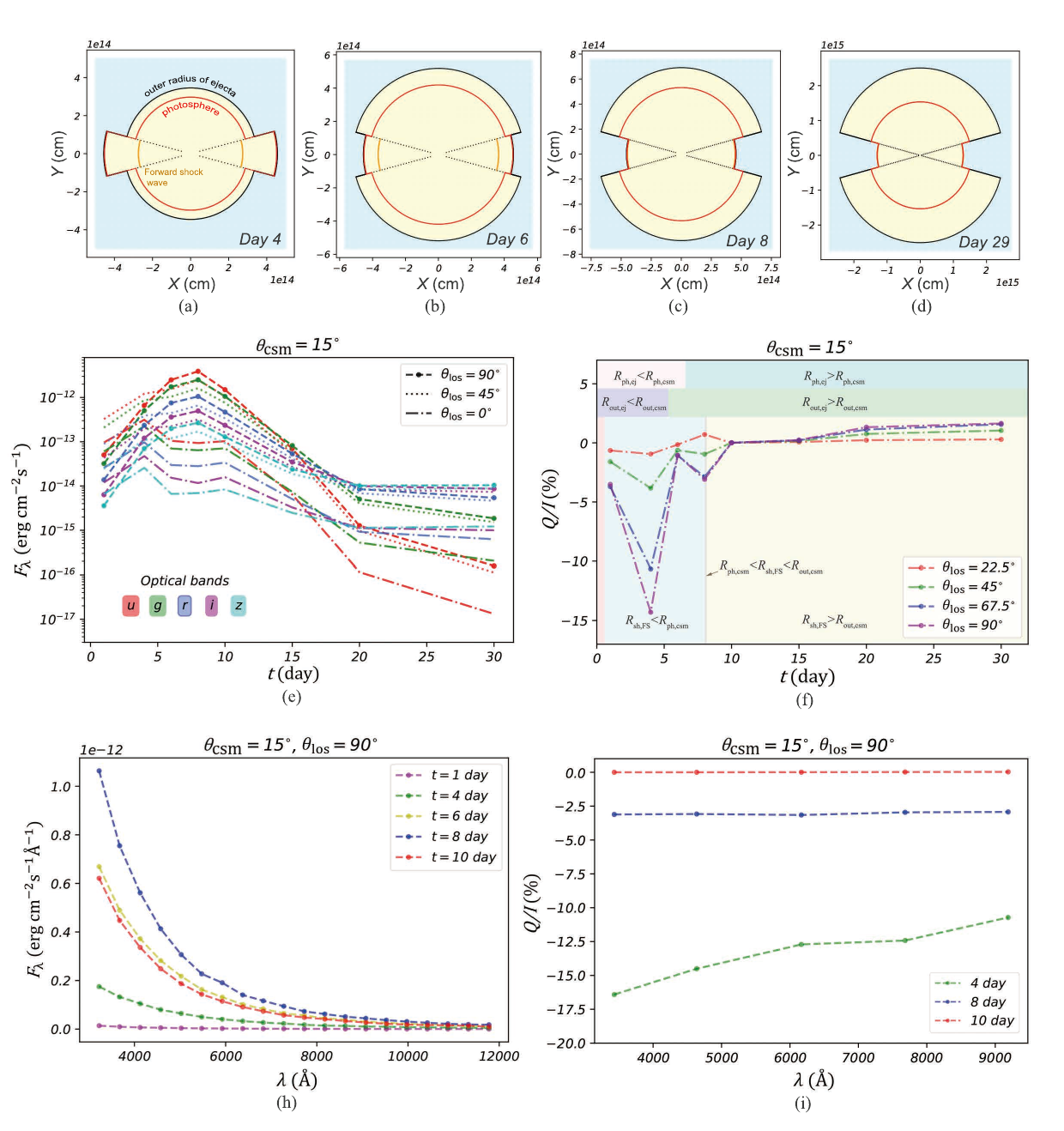}
\end{center}
\caption{
An example of the interaction between the expanding SN ejecta and a shell of CSM torus ($s=0$).
All phases are given in the rest frame and relative to the first light of the SN. As summarized by the schematics in the top row, panels (a)-(d) show the evolution of the outer radius of the ejecta (black solid line), the radius of the photosphere (red solid line), and the radius of the forward shock wave (yellow solid line) over time. 
Panel (e) presents the corresponding $ugriz$ spectral flux density per unit wavelength ($F_{\lambda}$) of the SN as seen from aspect angles of $\theta_{\rm los}=$90, 45, and 0 degrees as indicated by the legend.
Panel (f) displays the calculated time-evolution of the intensity-normalized polarization degree $Q/I$ for different viewing angles as marked by the legend.
The color-shaded areas indicate different stages of the interaction as characterized by the relative sizes among the photospheric radius in the ejecta and the CSM ($R_{\rm{ph,ej}}$,$R_{\rm{ph,CSM}}$), the outer radius of the ejecta and the CSM ($R_{\rm{out,ej}}$,$R_{\rm{out,CSM}}$), and the forward shock radius ($R_{\rm{sh,FS}}$). 
Panels (h) and (i) in the bottom row illustrate the wavelength dependence of the observed $F_{\lambda}$ and $Q/I$, respectively. Different viewing angles are indicated by the legend.}
\label{fig:Fig_n2}
\end{figure*}

\section{Result}

\begin{figure}[tbph]
\begin{center}
\includegraphics[width=0.49\textwidth,angle=0]{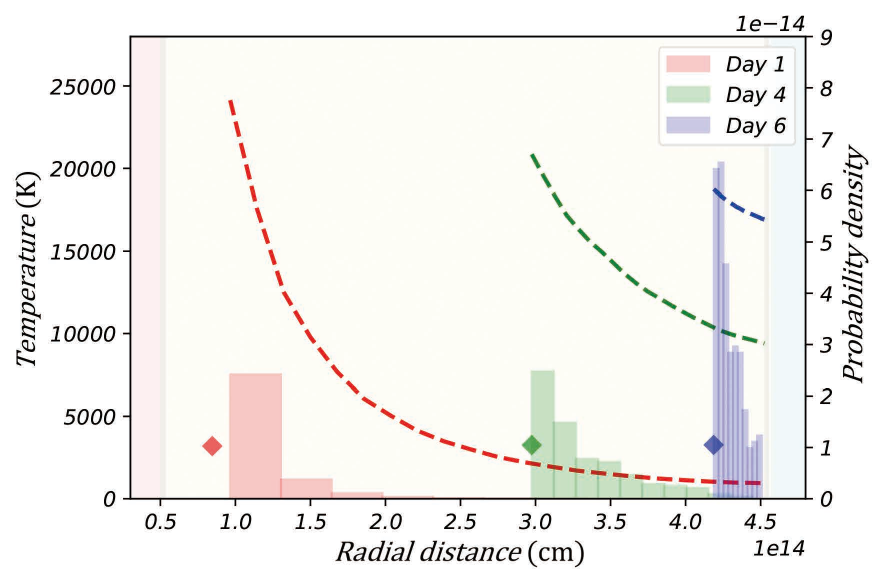}
\end{center}
\caption{Temperature structure as a function of radial distance of the CSM torus that lies above the photosphere at 1, 4, and 6 days after the first light, as shown by the red-, green-, and blue-dashed curves, respectively.
The red, green, and blue histograms present the corresponding probability density distribution of the photons along the radial direction, normalized to the total number of photons. 
The unit of the right-hand ordinate is photons per meter (photons/m). 
The probability density measured at each bin, scaled by a factor of '1e-14', can be read off the right-hand ordinates. The three diamonds represent the temperatures of the supernova ejecta photosphere in region \uppercase\expandafter{\romannumeral2}.
The yellow-shaded area represents the radial extension of the adopted configuration of the torus-like CSM. The red- and the blue-shaded areas indicate the ranges within the inner and beyond the outer boundaries of the CSM torus, respectively.
}
\label{fig:Fig_n3}
\end{figure}

\begin{figure*}[tbph]
\begin{center}
\includegraphics[width=1\textwidth,angle=0]{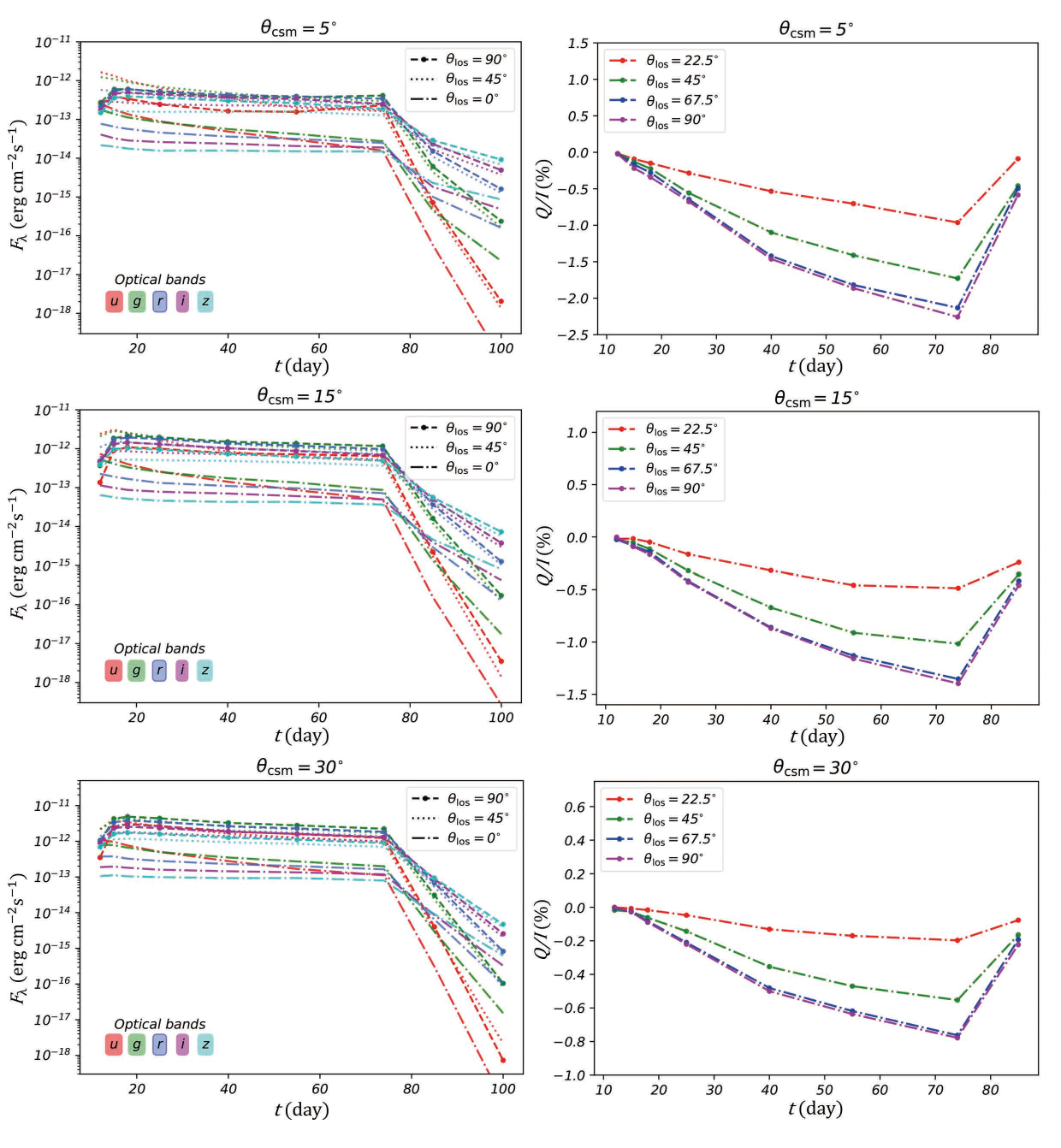}
\end{center}
\caption{An example of the interaction between the expanding SN ejecta and a CSM torus with a wind-like ($s=2$) density structure.
All phases are given in the rest frame and relative to the first light of the SN.
The left column shows, from top to bottom, the SN lightcurves ($F_{\lambda}$)
for $\theta_{\rm{csm}} = 5^{\circ}$,$15^{\circ}$,
and
$30^{\circ}$, respectively.
The right subpanels in each row display the corresponding $Q/I$. Viewing angles $\theta_{\rm los}$ are indicated by the legend in each subpanel.
}
\label{fig:Fig_n4}
\end{figure*}

\begin{figure}[tbph]
\begin{center}
\includegraphics[width=0.47\textwidth,angle=0]{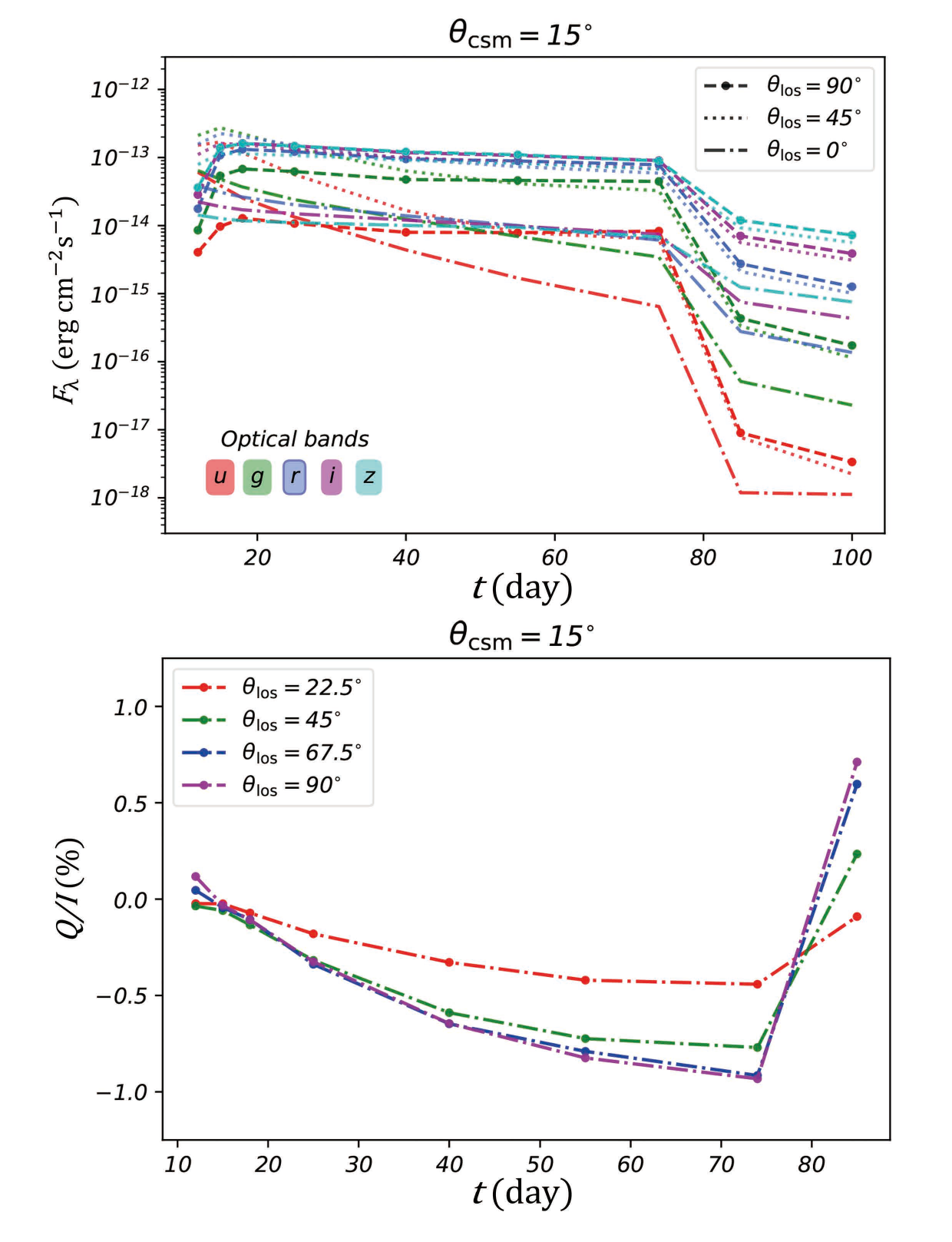}
\end{center}
\caption{Similar to Figure.~\ref{fig:Fig_n4} but for $\theta_{\rm{csm}} = 15^{\circ}$ and a lower kinetic-luminosity conversion efficiency (i.e., $\epsilon_{\rm{FS}}=\epsilon_{\rm{RS}}=0.1$).
}
\label{fig:Fig_n5}
\end{figure}

Our calculations suggest two main factors that affect the polarization signal when the SN ejecta interacts with a torus-like CSM.
The first one is the extra energy injected into the CSM torus by the FS and the RS waves during the interaction phase, which introduces a significant difference in the emitted radiations between regions I and II. The second one gives the deviation of the initially spherical SN ejecta caused by its interaction with the aspherical, torus-like CSM.
Parameters that characterize these two factors include the time-evolution of the luminosity difference between the SN ejecta and the CSM torus, $\theta_{\rm{csm}}$, the viewing angle ($\theta_{\rm los}$), and the time scale of the ejecta overriding the torus-like CSM.

\subsection{Ejecta interacting with immediate CSM Torus
%\textcolor{red}{YY: `with an immediate thin shell of CSM'}
}~\label{sec:csm_nearby}

When the SN ejecta interacts with an ambient thin torus of CSM,  whose radial density profile can be described by a uniform shell, namely $s=0$ in Equation~\ref{rho}, our calculations suggest that the interaction can produce a rather high level of continuum polarization.
In Figure~\ref{fig:Fig_n2} we present a case that adopts
$s=0$, $R_{\text{CSM,in}}=5 \times 10^{13}$ cm, $M_{\rm ej}=10 M_{\odot}$, $M_{\rm CSM}=0.1 M_{\odot}$, $E_{\rm {SN}}=10^{51}$ erg, $v_{\rm{SN}} =10^4 \text{ km s}^{-1}$, $\rho _{\text{CSM,in}}=5 \times 10^{-13} \text{g cm}^{-3}$, $\theta_{\rm{csm}}=15^{\circ}$, $\epsilon_{\rm{FS}}=\epsilon_{\rm{RS}}=0.8$ and $M_{\rm Ni}=0.032 M_{\odot}$. 
As illustrated in the top row of Figure~\ref{fig:Fig_n2}, in Region \uppercase\expandafter{\romannumeral1} (equatorial), the FS
propagates through the optically thick CSM in the eight days (Figures~\ref{fig:Fig_n2}a--\ref{fig:Fig_n2}c).
According to our calculations, within the first $\sim$8 days, the emission measured for region \uppercase\expandafter{\romannumeral1} is $\sim$3 order of magnitude stronger compared to that of region \uppercase\expandafter{\romannumeral2}, thus being the dominant radiation source during such an early interaction phase.

In the first few days after the SN explosion, due to the radial optical depth in the CSM disk being significantly greater than unity, most of the photons generated by the FS are emitted from the surface of the CSM disk near the shock front along the latitudinal direction. 
As the FS propagates outward, the diffusion time of photons in the radial direction of the CSM disk gradually decreases, leading to an increase in the photospheric luminosity at larger radial within the CSM disk.
The polarization level also gradually increases due to the brightening of the highly aspherical photosphere.
The high-luminosity photosphere of the CSM disk, exposed beyond the SN photosphere, can generate significant polarization signals (up to $\sim 15\%$) when both the latitudinal and radial surfaces of the CSM disk play prominent roles, as observed on day 4.
As the SN ejecta continues to expand, the overall shape of the photosphere gradually becomes spherical, leading to a progressively decreasing continuum polarization.
At around day 8 when the photons diffuse out the radial direction of the torus CSM, the lightcurve reaches its peak at $\theta_{\rm los} = 90^{\circ}$ (with $\theta_{\rm los} = 0^\circ$ directed towards the +z axis). 
The emission excess towards the equatorial plane would lead to rise of the continuum polarization, which becomes most prominent if seen from an equator-on perspective.
As the viewing perspective moves toward a pole-on scenario, photons emitted from region \uppercase\expandafter{\romannumeral2} become progressively obscured by region \uppercase\expandafter{\romannumeral1} (Figure~\ref{fig:Fig_n2}c).
resulting in a monotonically decrease of the total luminosity and a flip in the observed polarization position angle.
As the interaction process ends, the luminosity of region \uppercase\expandafter{\romannumeral1} gradually dies out and the steadily-expanding ejecta in region \uppercase\expandafter{\romannumeral2} would take over. The overall shape of the emitting zone will then gradually transfer from an oblate ($\lesssim$day 8) to a prolate ellipsoid because the ejecta goes faster at the poles. Such a geometric alternation would manifest as a flip of the sign of the polarization, namely from negative to positive as measured by $Q/I$ (Figure~\ref{fig:Fig_n2}f).

Another remarkable signature during the ejecta interacts with the CSM torus process
is that the polarization exhibits a wavelength dependence (see, Figure~\ref{fig:Fig_n2}i).
Typically, polarization aroused from Thomson scattering of free electrons with a globally aspherical distribution is considered wavelength independent.
Our calculation suggests that the polarization of the ejecta-CSM interaction exhibits a strong wavelength dependence, which
is mainly due to the interaction process induces
a significant temperature gradient along the radial direction of the disk disk. 
We illustrate this effect by calculating the radial structure of the temperature in Figure~\ref{fig:Fig_n3}.
Soon after the SN explosion, the majority of high-frequency photons are radiated from the latitudinal surface of the disk away from $R_{\rm{ph,CSM}}$, while a small fraction of low-frequency photons are radiated from the radial surface of the disk close to $R_{\rm{ph,CSM}}$.
As the shock front continues to propagate outward, the temperature difference of the CSM disk in the radial direction decreases and photons can gradually be radiated from the CSM in all directions, resulting in a wavelength-dependence that is no longer significant.

We note that different temperatures are also found in regions \uppercase\expandafter{\romannumeral1} ($\approx$3200\,K at day 4) and \uppercase\expandafter{\romannumeral2} ($\approx$21000\,K at day 4).
However, the significant luminosity disparity between the two regions suggests
that the low-frequency, low-energy photons emitted from region \uppercase\expandafter{\romannumeral1} have only a relatively weak effect on the wavelength dependence of the polarization, which can be considered negligible.
When the ejecta photosphere covers the latitudinal surface of region \uppercase\expandafter{\romannumeral1}, 
photons from the radial surface of the CSM disk and photons from the low-temperature region \uppercase\expandafter{\romannumeral1} create a small polarization difference at the red and blue ends (approximately $0.2\%$ on day 8).

\subsection{SN Ejecta Interacting with a Detached CSM Torus
}~\label{sec:csm_far}
We also considered the case of a radially extended torus of CSM located relatively far from the progenitor star of the supernova.
Following a similar procedure as detailed in Section~\ref{sec:csm_nearby}, we found that the presence of a detached CSM disk would produce overall lower levels of the continuum polarization that have similar trends to those of an immediate thin shell of the CSM.
As illustrated in Figure~\ref{fig:Fig_n4}, we present our calculation adopting the following parameters:
$s=2$, $R_{\text{CSM,in}} =1 \times 10^{15}$ cm, $M_{\rm ej}=10 M_{\odot}$, $M_{\rm CSM}=1 M_{\odot}$, $E_{\rm {SN}}=10^{51}$ erg, $v_{\rm{SN}} =10^4 \text{ km s}^{-1}$, $\rho _{\text{CSM,in}}=5 \times 10^{-14} \text{g cm}^{-3}$, $\epsilon_{\rm{FS}}=\epsilon_{\rm{RS}}=0.8$ and $M_{\rm Ni}=0.032 M_{\odot}$. 
In this example, region I undergoes a prolonged interaction process. As the luminosity from radioactive elements increases, the difference between the luminosities of Regions I and II decreases monotonically until day 74, leading to a reduced level of the continuum polarization.
The lightcurve enters a plateau after the peak until $R_{\rm{sh,FS}}=R_{\rm{ph,CSM}}$, 
corresponding to day 74 in Figure~\ref{fig:Fig_n5}.
After that, the lightcurve is mostly powered by the emission from the ejecta seen from region II.

We remark that since $R_{\rm{ph,ej}}$ remains smaller than $R_{\rm{ph,CSM}}$ in the CSM disk along the radial direction before the shock breakout occurs, there is no scenario where the photosphere of Region II obscures region I. Therefore, as the FS continuously propagates towards the disk's edge, the polarization degree gradually increases, until reaching its peak when the FS reaches the $R_{\rm{ph,CSM}}$.
As the luminosity of region I decreases after the shock breakout, the ejecta becomes more like a prolate ellipsoid because the ejecta expands faster at the poles. Such an oblate-to-prolate transformation of the overall geometry of the emitting zone thus leads to a sign flip of the integrated polarization (see the right column of Figure~\ref{fig:Fig_n4} after $\approx$day 75).
We also calculated the time-evolution of the continuum polarization for the cases of the CSM torus with a range of opening angles.
As $\theta_{\rm{csm}}$ increases, the observed polarization degree gradually decreases. For instance, when $\theta_{\rm{csm}} = 30^{\circ}$, the polarization level significantly drops to below 1\% (see the bottom row of Figure~\ref{fig:Fig_n4}).

Finally, following \cite{Nagao2020}, we examined the interaction process with a lower kinetic energy conversion efficiency, namely $\epsilon_{\rm{FS}}=\epsilon_{\rm{FS}}=0.1$
under the same configuration as the radially-extended envelope. The result for a half opening angle of $\theta_{\rm{csm}}=15^{\circ}$ is shown in Figure~\ref{fig:Fig_n5}.
During the interaction process until day 74, the luminosities of regions I and II are only differed by an order of magnitude, resulting in a polarization degree of less than 1\%.
Towards the end of the interaction process, the fast expanding ejecta near the poles would produce a prolate photosphere, thus producing a flip of the sign of the continuum polarization (see the lower panel of Figure~\ref{fig:Fig_n5}).

\section{Conclusion and Discussion} \label{sec:Conclusion and Discussion}

In this letter, we investigate the temporal evolution of the polarization of SN interacting with various configurations of dense, torus-like CSM.
Our calculations suggest that the integrated continuum polarization would result from an interplay between the overall geometry and the non-uniformly distributed emitting regions, both are time-variant.
When the SN ejecta interact with the CSM disk, the CSM disk exposed outside the ejecta photosphere leads to an overall oblate ellipsoidal shape of the emitting zone.
Additionally, the increased difference of the emitted flux between the CSM torus (region I) and the polar regions (region II) within the first 4 days after the explosion would produce a further elevated level of the continuum polarization (Figures~\ref{fig:Fig_n1}a and ~\ref{fig:Fig_n1}f).
This effect is particularly pronounced in the case of a shell-like CSM located in close vicinity of the progenitor star.
The estimated level of the peak polarization can be high as $\approx$15\% (Figure~\ref{fig:Fig_n2}f), which may account for the observed record-breaking high polarization a few days after the transient explosion, e.g., the $\sim$7\% polarization in blue wavelength range within $\sim$5 days of the explosion of AT\,2018cow \citep{Maund2023}.
As the ejecta expands faster in polar regions (II), it will progressively cover the emitting regions along the CSM torus (Figure~\ref{fig:Fig_n2}b). As a consequence, photons produced by the ejecta-CSM interaction will predominantly diffuse out along the radial direction of the CSM torus.
At this point, the continuum polarization would result in an interplay between the geometry of the emitting zone and the luminosity difference between the pole ejecta (region II) and the radial surface of the CSM torus.
Especially in cases where the radial luminosity of the CSM is more dominant, a secondary peak of the polarization might be observed (see, e.g., Figure~\ref{fig:Fig_n2}f at $\approx$day 8).
A reversal of the sign of the polarization may occur after the secondary polarization peak, followed by a steady increase in the polarization level. The latter can be understood as the ejecta's expansion towards the poles, which would cause an increased deviation from spherical symmetry.
Some dynamical simulations show that as the ejecta expands, it will gradually engulf the CSM disk \citep{McDowell2018}. The polarization at this stage may be determined by a rather complicated process that carves
the post engulfment geometry, which is beyond the scope of this letter.

We found that when the SN ejecta interacts with a CSM torus, the non-uniform temperature distribution between the polar and equatorial regions would produce a wavelength-dependent polarization. This can be tested by multi-band imaging polarimetry or spectropolarimetry.
In previous studies, wavelength-dependent polarization has typically been attributed to wavelength-dependent line opacity \citep{Howell2001,Maund2013,Inserra2016,Yang2020}.
In the first few days after the SN explosion, as a consequence of the ejecta-CSM interaction, the temperature would exhibits a dramatic decrease towards larger radius of the CSM torus (Figure~\ref{fig:Fig_n3}).
Photons emitted from the hotter inner regions of the disk 
(region I) dominate the polarization at shorter wavelengths, while a small fraction of photons emitted from the cooler outer regions manipulate
the polarization at longer wavelengths.
The latter is also affected by the photons emitted from the relatively cooler region II. However, such radiation from the polar ejecta remains approximately more than one order of magnitude lower compared to that from the outer part of the region I.

Despite our calculations predict the temporal evolution of the polarization for both a thin shell and a radially extended torus of CSM, we remark that the detailed process of interaction can be affected by various other factors.
For instance, multiple components of CSM may form a rather complicated density structure and thus may manifest as time-variant polarization that differs from our calculations \citep{liu2017}.
The early-time evolution of the broad-band polarization measured for the fast blue optical transient AT\,2018cow exhibits two peaks at $\approx$days 5.7 and 13, reaching $p\sim7\%$ in $r$ and $p\sim2\%$ in $b$, respectively \citep{Maund2023}.
Such behavior is broadly consistent with that predicted for the SN ejecta interacting with a thin shell of CSM torus.
For the CSM confined within a rather small distance to the progenitor star, the corresponding timescale of the interaction will also be compressed.
For instance, an AT2018cow-like event with a compact $s=0$ CSM disk may require polarization diagnostics to be carried out within $\sim$day after the explosion. Additionally, hours-cadence sampling would be necessary to resolve the first and the secondary peaks in their polarization time evolution.
Future deployment of high-precision polarimeters with rapid-response capability will enable unique measurements of the geometry of the ejecta-CSM interaction process over a broad range of timescales. 
The polarization time sequence for various types of transients will set constraints on the evolution and final mass loss of their progenitors.

\begin{acknowledgments}
We thank Bing Zhang, Lifan Wang, Maokai Hu and Justyn R Maund for their valuable discussions, and the anonymous reviewers for their helpful comments that improved the manuscript.
Xudong expresses his heartfelt gratitude to his grandmother in heaven for her love and encouragement during her lifetime. 
This work is supported by the National Natural Science Foundation of China (Projects 12373040,12021003), the National SKA Program of China (2022SKA0130100) and the Fundamental Research Funds for the Central Universities.
\end{acknowledgments}

\bibliographystyle{aasjournal}

\begin{thebibliography}{}


\bibitem[Andrews et al.(2017)]{Andrews2017} Andrews, J. E., Smith, N., McCully, C., et al. 2017, MNRAS, 471, 4047, doi:10.1093/mnras/stx1844

\bibitem[Arnett(1982)]{Arnett82} Arnett, W.~D.\ 1982, \apj, 253, 785, doi: 10.1086/159681

\bibitem[Bellm et al.(2019)]{Bellm2019} Bellm, E. C., Kulkarni, S. R., Graham, M. J., et al. 2019, PASP, 131, 018002, doi: 10.1088/1538-3873/aaecbe

\bibitem[Bruch et al.(2021)]{Bruch2021} Bruch, R. J., Gal-Yam, A., Schulze, S., et al. 2021, ApJ, 912, 46, doi: 10.3847/1538-4357/abef05

\bibitem[Bruch et al.(2023)]{Bruch2023} Bruch, R. J., Gal-Yam, A., Yaron, O., et al. 2023, ApJ, 952, 119, doi: 10.3847/1538-4357/acd8be

\bibitem[Bulla et al.(2015)]{Bulla15} Bulla, M., Sim, S. A., \& Kromer, M. 2015, \mnras, 450, 967, doi: 10.1093/mnras/stv657

\bibitem[Bulla(2017)]{Bulla2017} Bulla, M.\ 2017, PhD thesis, Astrophysics Research Centre, School of Mathematics and Physics, Queen’s University Belfast, Belfast BT71NN, UK

\bibitem[Bilinski et al.(2023)]{Bilinski2023} Bilinski, C., Smith, N., Williams, G. G., et al. 2023, \mnras, 529, 2, doi:10.1093/mnras/stae380

\bibitem[Chandrasekhar(1960)]{chandrasekhar1960} Chandrasekhar, S.\ 1960, Radiative Transfer. Dover Press, New York

\bibitem[Chatzopoulos, Wheeler \& Vinko(2009)]{Chatzopoulos09} Chatzopoulos, E., Wheeler, J. C., \& Vinko, J. 2009, \apj, 704, 1251

\bibitem[Chatzopoulos et al.(2012)]{Chatzopoulos12} Chatzopoulos E., Wheeler J. C., Vinko J., 2012, ApJ, 746, 121, doi: 10.1088/0004-637X/746/2/121

\bibitem[Chatzopoulos et al.(2013)]{Chatzopoulos13}Chatzopoulos, E., Wheeler, J. C., Vinko, J., Horvath, Z. L., \& Nagy, A. 2013, ApJ, 773, 76, doi: 10.1088/0004-637X/773/1/76


\bibitem[Chevalier(1982)]{Chev1982} Chevalier, R.~A.\ 1982, \apj, 258, 790, doi: 10.1086/160126

\bibitem[Chevalier \& Soker(1989)]{chevalier89} Chevalier, R.~A. \& Soker, N.\ 1989, \apj, 341, 867, doi: 10.1086/167545

\bibitem[Chevalier \& Fransson(1994)]{Chevalier1994} Chevalier, R. A., \& Fransson, C. 1994, ApJ, 420, 268, doi: 10.1086/173557

\bibitem[Chevalier(2012)]{Chevalier12}Chevalier R. A., 2012, \apjl, 752, L2, doi: 10.1088/2041-8205/752/1/L2

\bibitem[Chevalier \& Fransson(2017)]{Chevalier2017} Chevalier, R. A., \& Fransson, C. 2017, Thermal and Non-thermal Emission from Circumstellar Interaction, ed. A. W. Alsabti \& P. Murdin, 875, doi: 10.1007/978-3-319-21846-5$\_$34

\bibitem[Chevalier \& Irwin(2011)]{Chevalier2011} Chevalier, R. A., \& Irwin, C. M. 2011, ApJL, 729, L6, doi: 10.1088/2041-8205/729/1/L6

\bibitem[Clocchiatti \& Wheeler(1997)]{Clo1997} Clocchiatti, A., \& Wheeler, J. C. 1997, \apj, 491, 375, doi: 10.1086/304961

\bibitem[Code \& Whitney(1995)]{Code95} Code A. D., Whitney B. A.\ 1995, \apj, 441, 400, doi: Polarization from scattering in blobs

\bibitem[Dessart et al.(2015)]{Dessart2015} Dessart, L., Audit, E., \& Hillier, D. J. 2015, MNRAS, 449, 4304, doi: 10.1093/mnras/stv609

\bibitem[Drout et al.(2014)]{Drout2014} Drout, M. R., Chornock, R., Soderberg, A. M., et al. 2014, \apj, 794, 23, doi: 10.1088/0004-637X/794/1/23

\bibitem[Fuller(2017)]{Fuller2017} Fuller J., 2017, \mnras, 470, 1642, doi:10.1093/mnras/stx1314

\bibitem[Gal-Yam(2012)]{Gal-Yam2012} Gal-Yam, A. 2012, Science, 337, 927, doi: 10.1126/science.1203601

\bibitem[Gal-Yam et al.(2014)]{Gal-Yam2014}Gal-Yam, A., Arcavi, I., Ofek, E. O., et al. 2014, Nature, 509, 471, doi: 10.1038/nature13304

\bibitem[Ginzburg \& Balberg(2012)]{Ginzburg2012} Ginzburg, S., \& Balberg, S. 2012, ApJ, 757, 178, doi: 10.1088/0004-637X/757/2/178

\bibitem[Graham et al.(2014)]{Graham2014} Graham, M. J., Kulkarni, S. R., Bellm, E. C., et al. 2019, PASP, 131, 078001, doi: 10.1088/1538-3873/ab006cH

\bibitem[Ho et al.(2019)]{Ho2019} Ho, A. Y. Q., Goldstein, D. A., Schulze, S., et al. 2019, \apj, 887, 169, doi: 10.3847/1538-4357/ab55ec

\bibitem[H{\"o}flich(1991)]{Hoflich91} H{\"o}flich, P.\ 1991, \aap, 246, 481

\bibitem[Howell et al.(2001)]{Howell2001} Howell, D. A., Hoeflich, P., Wang, L., \& Wheeler, J. C.
2001, ApJ, 556, 302, doi: 10.1086/321584

\bibitem[Insrra et al.(2013)]{Inserra2013} Inserra, C., Smartt, S. J., Jerkstrand, A., et al. 2013, \apj, 770, 128, doi: 10.1088/0004-637X/770/2/128

\bibitem[Inserra et al.(2016)]{Inserra2016} Inserra, C., Bulla, M., Sim, S. A., \& Smartt, S. J. 2016,
ApJ, 831, 79, doi: 10.3847/0004-637X/831/1/79

\bibitem[Inserra et al.(2018)]{Inserra2018} Inserra, C., Smartt, S. J., Gall, E. E. E., et al. 2018, MNRAS, 475, 1046, doi: 10.1093/mnras/stx3179

\bibitem[Kasen(2003)]{Kasen03} Kasen, D., et al.\ 2003, \apj, 593, 788, doi: 10.1086/376601

\bibitem[Kasen et al.(2016)]{kasen16} Kasen, D., Metzger, B.~D., \& Bildsten, L.\ 2016, \apj, 821, 36, doi: 10.3847/0004-637X/821/1/36


\bibitem[Khatami \& Kasen(2019)]{Khatami2019} Khatami, D. K., \& Kasen, D. N.\ 2019, \apj, 878, 56, doi: 10.3847/1538-4357/ab1f09

\bibitem[Kurf{\"u}rst \& Krti{\v{c}}ka(2019)]{Kurfurst2019} Kurf{\"u}rst P., Krti{\v{c}}ka J., 2019, A\&A, 625, A24, doi: 10.1051/0004-6361/201833429

\bibitem[Liu et al.(2017)]{liu2017}Liu, L.-D., Wang, L.-J., Wang, S.-Q., \& Dai, Z.-G. 2018, \apj, 856, 59, doi: 10.3847/1538-4357/aab157

\bibitem[Lucy(1999)]{Lucy99} Lucy, L. B.\ 1999, \aap, 345, 211

\bibitem[Margutti et al.(2019)]{Margutti2019} Margutti, R., Metzger, B. D., Chornock, R., et al. 2019, \apj, 872, 18, doi: 10.3847/1538-4357/aafa01

\bibitem[Margalit \& Ho(2022)]{Margutti2022} Margalit, B., Quataert, E., \& Ho, A. Y. Q. 2022, ApJ, 928, 122, doi: 10.3847/1538-4357/ac53b0

\bibitem[Matzner \& McKee(1999)]{matzner99} Matzner, C.~D. \& McKee, C.~F.\ 1999, \apj, 510, 379, doi: 10.1086/306571

\bibitem[Mazzali \& Lucy(1993)]{Mazzali93} Mazzali, P. A., \& Lucy, L. B. 1993, \aap, 279, 447

\bibitem[Metzger \& Pejcha(2017)]{Metzger2017} Metzger, B. D. \& Pejcha, O. 2017, \mnras, 471, 3200, doi: 10.1093/mnras/stx1768

\bibitem[Mauerhan et al.(2024)]{Mauerhan2024} Mauerhan J. C., Smith N., Williams G. G., Smith P. S., Filippenko A. V., Bilinski C., Zheng W., et al., 2024, \mnras, 527, 6090, doi: 10.1093/mnras/stad3579

\bibitem[Maund et al.(2013)]{Maund2013} Maund, J. R., Spyromilio, J., Hoflich, P. A., et al. 2013,
MNRAS, 433, L20, doi: 10.1093/mnrasl/slt050

\bibitem[Maund et al.(2023)]{Maund2023} Maund, J. R., H{\"o}flich, P. A., Steele, I. A., et al. 2023,
MNRAS, 521, 3323, doi: 10.1093/mnras/stad539

\bibitem[McDowell, Duffell \& Kasen(2018)]{McDowell2018} McDowell A.~T., Duffell P.~C., Kasen D., 2018, ApJ, 856, 29, doi: 10.3847/1538-4357/aaa96e

\bibitem[Metzger(2022)]{Metzger2022} Metzger, B. D. 2022, \apj, 932, 84, doi: 10.3847/1538-4357/ac6d59

\bibitem[Mohamed \& Podsiadlowski (2007)]{Mohamed2007} Mohamed, S., \& Podsiadlowski, P. 2007, in Astronomical Society of the Pacific Conference Series, Vol. 372, 15th European Workshop on White Dwarfs, ed. R. Napiwotzki \& M. R. Burleigh, 397

\bibitem[Morozova et al.(2017)]{Morozova2017} Morozova, V., Piro, A. L., \& Valenti, S. 2017, \apj, 838, 28, doi: 10.3847/1538-4357/aa6251

\bibitem[Moriya et al.(2013)]{Moriya2013} Moriya, T. J., Blinnikov, S. I., Tominaga, N., et al. 2013, MNRAS, 428, 1020, doi:10.1093/mnras/sts075

\bibitem[Nagao,  Maeda \& Ouchi(2020)]{Nagao2020} Nagao, T., Maeda, K., \& Ouchi, R. 2020, MNRAS, 497, 5395, doi:10.1093/mnras/staa2360

\bibitem[Ofek et al.(2010)]{Ofek2010} Ofek, E. O., Rabinak, I., Neill, J. D., et al. 2010, ApJ, 724, 1396, doi: 10.1088/0004-637X/724/2/1396

\bibitem[Pejcha et al.(2016)]{Pejcha2016} Pejcha, O., Metzger, B. D., \& Tomida, K. 2016, \mnras, 461, 2527, doi: 10.1093/mnras/stw1481

\bibitem[Pursiainen et al.(2022)]{Pursiainen2022} Pursiainen, M., Leloudas, G., Paraskeva, E., et al. 2022,
A\&A, 666, A30, doi: 10.1051/0004-6361/202243256 

\bibitem[Quimby et al.(2007)]{Quimby2007} Quimby, R. M., Wheeler, J. C., H{\"o}flich, P., et al. 2007, \apj, 666, 1093, doi: 10.1086/520532

\bibitem[Quataert \& Shiode(2012)]{Quataert2012}Quataert E., Shiode J., 2012, \mnras, 423, L92, doi: 10.1111/j.1745-3933.2012.01264.x 

\bibitem[Shapiro \& Sutherland(1982)]{Shapiro82} Shapiro, P. R., \& Sutherland, P. G.\ 1982, \apj, 263, 902, doi: 10.1086/160559

\bibitem[Smith et al.(2007)]{Smith2007} Smith, N., Li, W., Foley, R. J., et al. 2007, \apj, 666, 1116, doi: 10.1086/519949

\bibitem[Soker \& Kashi (2013)]{Soker13}Soker N., Kashi A., 2013, \apjl, 764, L6, doi: 10.1088/2041-8205/764/1/L6

\bibitem[Suzuki et al.(2019)]{Suzuki2019} Suzuki, A., Moriya, T. J., \& Takiwaki, T. 2019, ApJ, 887, 249, doi: 10.3847/1538-4357/ab5a83

\bibitem[Suzuki et al.(2020)]{Suzuki2020} Suzuki, A., Moriya, T. J., \& Takiwaki, T. 2020, \apj, 899, 56, doi: 10.3847/1538-4357/aba0ba

\bibitem[Uno et al.(2023)]{Uno2023} Uno K., Nagao T., Maeda K., Kuncarayakti H., Tanaka M., Kawabata K. S., Nakaoka T., et al., 2023, \apj, 944, 2, doi: 10.3847/1538-4357/acb5eb

\bibitem[Valenti et al.(2008)]{Valenti2008} Valenti, S., Benetti, S., Cappellaro, E., et al. 2008, \mnras, 383, 1485, doi: 10.1111/j.1365-2966.2007.12647.x

\bibitem[Wen et al.(2023)]{Wen2023} Wen, X.-D., Gao, H., Ai, S.-K., et al. 2023, \apj, 955, 9, doi: 10.3847/1538-4357/acef11

\bibitem[Wang et al.(2017)]{Wang2017} Wang, S. Q., Wang, L. J., Dai, Z. G., et al. 2015, \apj, 807, 147, doi: 10.1088/0004-637X/807/2/147

\bibitem[Wang et al.(2019)]{Wang2019} Wang, L. J., Wang, X. F., Cano, Z., et al. 2019, \mnras, 489, 1110, doi: 10.1093/mnras/stz2184

\bibitem[Wang \& Wheeler(2008)]{Wang08} Wang, L., \& Wheeler, J.~C.\ 2008, \araa, 46, 433, doi: 10.1146/annurev.astro.46.060407.145139

\bibitem[Whitney(2011)]{Whitney2011} Whitney B. A.\ 2011, Bulletin of the Astronomical Society of India, 39, 101, doi: 10.48550/arXiv.1104.4990

\bibitem[Vasylyev et al.(2023)]{Vasylyev2023} Vasylyev, S. S., Yang, Y., Filippenko, A. V., et al. 2023, ApJL, 955, L37, doi:  
10.3847/2041-8213/acf1a3

\bibitem[Vasylyev et al.(2024)]{Vasylyev2024} Vasylyev, S. S., Yang, Y., Patra, K. C., et al. 2024, MNRAS, 527, 3106, doi: 10.1093/mnras/stad3352

\bibitem[Yaron et al.(2017)]{Yaron2017} Yaron, O., Perley, D., Gal-Yam, A., et al. 2017, Nature Physics, 13, 510, doi: 10.1038/nphys4025

\bibitem[Yan et al.(2017)]{Yan2017} Yan, L., Lunnan, R., Perley, D. A., et al. 2017, ApJ, 848, 6, doi:  
10.3847/1538-4357/aa8993

\bibitem[Yang et al.(2020)]{Yang2020} Yang, Y., Hoeflich, P., Baade, D., et al. 2020, ApJ, 902, 46,
doi: 10.3847/1538-4357/aba759

\bibitem[Yang et al.(2023)]{Yang2023} Yang, Y., Baade, D., Hoeflich, P., et al. 2023, MNRAS, 519, 1618, doi: 10.1093/mnras/stac3477

\bibitem[Yoon \& Cantiello(2010)]{Yoon2010} Yoon S.-C., Cantiello M., 2010, \apjl, 717, L62, doi: 10.1088/2041-8205/717/1/L62

\end{thebibliography}

\end{document}